\documentclass[a4paper,11pt]{article}

\usepackage{jcappub}
\usepackage[T1]{fontenc}
\usepackage{tabularx}

\let\log\relax
\DeclareMathOperator{\log}{Log}

\def\bea{\begin{eqnarray}}
\def\eea{\end{eqnarray}}
\def\be{\begin{equation}}
\def\ee{\end{equation}}
\def\fr{\frac}

\def\la{\label}
\def\be{\begin{equation}}
\def\ee{\end{equation}}

\def\le{\left}
\def\ri{\right}

\title{\boldmath  Bound Dark Matter (BDM) towards solving the Small Scale Structure Problem}

\author[a]{Jorge Mastache}
\affiliation[a]{CONACYT-Mesoamerican Centre for Theoretical Physics, Universidad Aut\'{o}noma de Chiapas, \\ Carretera Zapata Km. 4, Real del Bosque, 29040, Tuxtla Guti\'{e}rrez, Chiapas, M\'{e}xico.}
\emailAdd{jhmastache@mctp.mx}

\author[b,c]{and Axel de la Macorra}
\affiliation[b]{Instituto de Fisica, Universidad Nacional Autonoma de Mexico, Apdo. Postal 20-364, 01000, CDMX, Mexico}
\affiliation[c]{ICC, University of Barcelona, Marti i Franques, 1, E08028 Barcelona, Spain.}
\emailAdd{macorra@fisica.unam.mx}

\abstract{
Cosmological observations such as structure formation, cosmic microwave background, and cosmic distance ladder set tight constraints to the amount and nature of dark matter (DM). In particular structure formation strongly constraints not only the amount of energy density but also the time when DM became non-relativistic, $a_{nr}$. Standard cold and thermic warm DM particles have a smooth transition from being relativistic at high energies to a non-relativistic regime since the mass of these particles is constant while the velocity redshifts with the expansion of the universe. However, here we explore the possibility that the DM particle acquires a non-perturbative mass at a phase transition scale and a scale factor $a_c$, e.g.  the mass of protons and neutrons is due to the binding energy of QCD through a non-perturbative process.  This transition acquired a more fundamental meaning for the Bound Dark Matter (BDM) model because they describe a particle getting its mass through a non-perturbative process. These BDM particles may go from being relativistic to non-relativistic at  $a_c$, which implies an abrupt transition of the velocity of the particles $v_{c}$ at that time, affecting  the equation of state and the cosmological evolution.  Here we  study  the cosmological impact of  the values of   $v_c$  and the scale transition $a_c$ and they  by reducing the free-streaming scale and therefore the small scale structure. Using CMB Plank, Supernovae SNIa, and baryon acoustic oscillation data, we constrain the valid region in the parameter space $a_c - v_c$ putting  upper bounds to $a_c$ but not restricting $v_c$. For instance, the transition must be $a_c < 2.66 \times 10^{-6}$ for  $v_c = 0$. We also find that the free-streaming and the Jeans mass of the dark matter particle is highly influenced by the velocity $v_c$, for example, a 3 keV WDM have a Jeans mass of $M_{fs} = 1.97 \times 10^7 {\rm M_\odot/h^3}$ but an equivalent BDM with the same $a_{nr}$ but an abrupt transition in the velocity, $v_c \sim 0$ would have a Jeans mass of $M_{fs} = 2.08\times10^4  {\rm M_\odot/h^3}$ which significantly changes the large scale structure formation.
}

\begin{document}

\maketitle
\flushbottom

\section{Introduction}\label{sec:introduction}

The measurement of the Cosmic Microwave Background (CMB) anisotropies \cite{Aghanim:2018eyx} and the mapping of the large-scale structures (LSS), through galaxy redshift surveys \cite{Abbott:2016ktf, Percival:2006gt} and type Ia SuperNovae \cite{Betoule:2014frx} have a great impact in our knowledge of the Universe. In particular, they have established the standard model for cosmology, the $\Lambda$CDM model, in which the content of the Universe consists of 68\% dark energy driving the expansion of the Universe,  4\% baryons and 28\% dark matter (DM)  at present time.  The nature of these two dark components, which account for up to $96\%$ of the energy content of the Universe,  is still  unknown. DM clustering properties have a major impact on structure formation and in this work, we explore the parameters that could constrain the essence of dark matter and help to solve the small scale crisis in the Universe.

A large number of candidates have been proposed for DM \cite{Bertone:2004pz} of which cold dark matter (CDM) has been the most popular. The CDM model has been successful in explaining large scale structure formation in the early Universe as well as abundances of galaxy clusters \cite{Springel:2005nw, Cooray:2002dia}. However, the clustering properties of structure at scales of the order of galaxy scales is not well understood. For instance, the number of satellite galaxies in our Galaxy is smaller than the expected from  $\Lambda$CDM model \cite{Kazantzidis:2003hb}, the so-called missing satellite problem. In a CDM scenario  the DM particles become non-relativistic at very early stages of the evolution of the Universe (e.g. for a masses larger than $\mathcal{O}$(MeV) for thermally produced particles)  and form structure at all relevant scales, small halos merge to form larger ones, a process that spans a wide of  range of scales, from galaxy clusters down to micro-haloes with masses down to the Earth's mass. However, this standard scenario seems to disagree with a number of observations. First,  the number of sub-haloes around a typical Milky Way galaxy, as identified by satellite galaxies, is an order of magnitude smaller than predicted by $\Lambda$CDM model  \cite{Kazantzidis:2003hb, BoylanKolchin:2011de}. 

At the same time, $\Lambda$CDM  also predicts steeply cuspy density profiles, causing a large fraction of haloes to survive as substructure inside larger haloes \cite{Navarro:1996gj}. Observed rotation curves for dwarf spheroidal dSph and low surface brightness (LSB) galaxies seem to indicate that their DM haloes prefer constant density cores \cite{Moore:1994yx} instead of steep cusps as predicted by the Navarro-Frank-White profile \cite{Navarro:1996gj}. LSB galaxies are diffuse, low luminosity systems, which kinematic is believed to be dominated by their host DM halos \cite{deBlok:2001hbg}. Assuming that LSB galaxies are in virial equilibrium, the stars act as tracers of the gravitational potential, therefore, they can be used as a probe of the DM density profile. Much better fits to dSph and LSB observations are found when using a cored halo model \cite{deBlok:2001hbg}. Cored halos have a mass-density that remains at an approximately constant value towards the center.
 
Many solutions to both of these problems have been proposed, and there are two main branches. First, the solution through baryon physics - star formation and halo evolution in the galaxy may be suppressed due to some baryonic process \cite{Penarrubia:2012bb}. Second, the DM solution in which the number of satellite galaxies is suppressed due to the kinematic properties of the DM particles \cite{Bode:2000gq}. The discussion is still in progress. Here we take the second branch by constraining the properties of DM particles that influence structure formation. 
 
Regarding the cosmological impact of DM particles, perhaps the two most important quantities to determine the cosmological effects, besides the amount of energy density, $\Omega_{\rm dm}$,  is the time when the DM particles became non-relativistic, defined by the scale factor $a_{nr}$, and the velocity dispersion of these particles at that time, $v_{nr}$.  For thermally decoupled warm dark matter (WDM) particles, we compute a one to one relationship between the WDM mass $m_{wdm}$ and $a_{nr}$. It is common to define the time when a particle became non-relativistic when the  momentum equals its mass, i.e. $p^2=m^2$,  which correspond to a velocity  $v_{nr}= 1/\sqrt{2}$  \cite{Bode:2000gq}. For example, Ly-$\alpha$ forest observations \cite{Viel:2013apy}, which measure clouds of neutral Hydrogen in the Universe at high redshift ($z \approx \mathcal{O}(1-3)$) set a  a lower limit  $m_{wdm} \gtrsim $ 3keV. Despite hydrogen gas may not accurately trace the distribution of dark matter and the lower limit for the mass is still unsettled (see more details in \cite{Hui:2016ltb}), we use a $m_{\rm wdm}=3$ keV WDM mass as a reference and compare it to our BDM model. A  $m_{\rm wdm}=3$ keV particle becames non-relativistic at $a_{nr}=3.18\times 10^{-8}$. Furthermore, it is usually assumed  a smooth evolution for the  velocity dispersion $v(a)$ for a thermal WDM. However, this assumption is based on the assumption that the mass of the particle is constant, which is not necessarily true in all cases. We could have DM with an abrupt evolution of $v(a)$ if the mass of the DM particle is due to the binding energy of elementary particles, for example the mass of neutrons or protons in the SM of particles, or in the Bound Dark Matter (BDM) model \cite{delaMacorra:2009yb} analyzed here. 

The BDM model, motivated by particle physics, assumes that elementary particles contained in a gauge group are nearly massless at high energies, but once the energy decreases the gauge force becomes strong (at a  scale factor defined by $a_c$ and energy $\Lambda_c$)  and forms neutral bound states which  acquire a non-perturbative mass proportional to $\Lambda_c$. The BDM particles are by hypothesis, not contained in the  SM of particle physics. Indeed, at high energies, the elementary particles (quarks) of the  Standard Model (SM) are weakly coupled however the strength of the gauge coupling constant increases for lower energies and eventually  becomes strong at the condensation energy scale $\Lambda_{\rm QCD}$ and scale factor denoted by $a_{\rm QCD}$.  At this scale, gauge-invariant states are created forming gauge neutral composite particles,  mesons (e.g. pions $\pi$) and baryons   (e.g. protons and neutrons),  at  $\Lambda_{\rm QCD}= (210 \pm 14)$MeV \cite{Tanabashi:2018oca},  with non  perturbative masses generated and being proportional to   $\Lambda_{\rm QCD}$ (e.g. $m_\pi \simeq 140$MeV,  $m_{n}\simeq 940$MeV) much larger the quark masses ($m_u\simeq 2.3$MeV,  $m_d\simeq  4.8$MeV). The mass of  the bound states (mesons and baryons) is due to the underlying gauge force and is independent of the bare mass of the original quarks. Since the mass of the bound states is much larger than the mass of the elementary particles the resulting velocity dispersion,  $v_c$, of these bound states is significantly reduced to the velocity  of the original elementary particles. Therefore, we expect to have an abrupt transition  for the velocity dispersion  $v_c$ at $a_c$  with   $v_c\ll v_{nr}\equiv1/\sqrt{2}$.
 
As in QCD, where the transition from fundamental elementary particles (quarks) to bound states (mesons and baryons) takes place from high to low the energies, our BDM model also forms bound states at lower energies. The transition from high to low energy densities can be encountered in different cosmological scenarios.  One case is as a consequence  of the expansion of the Universe, as it grows the energy density dilutes, and secondly  inside massive structures (e.g. galaxies) where  the energy density  increases with decreasing radius. Therefore, our BDM could help to ameliorate two of the main  CDM problems, namely  the missing satellite problem  and the cuspy energy density profiles in low-density galaxies \cite{delaMacorra:2011df}. Indeed,  the free streaming of the BDM particles prevents small halos and will  also have an  impact in the centre region of galaxies rendering a core galactic  profile.

Here we study the cosmological properties of BDM particles by constraining these two parameters, the scale factor $a_c$ and the velocity dispersion $v_c$, when the bound states acquire their non-perturbative mass, and by taking $a_c=a_{nr}$ and $v_c=v_{nr}=1/\sqrt{2}$ we recover WDM scenario as a limiting case of BDM.  We find that a BDM model that becomes non-relativistic at a scale factor 10 times larger than in a WDM model can have an equivalent free streaming scale $\lambda_{fs}$ and $k_{1/2}$, the mode where the power spectrum is suppressed by $50\%$ with regard to $\Lambda$CDM model, therefore rendering an equivalent suppression on small scale structures as WDM. Moreover, the Jeans mass can be reduced 2 orders of magnitude between a BDM model with a velocity $v_c = 1/\sqrt{2}$ (resembling WDM smooth transition) and a BDM with an abrupt transition, $v_c \sim 0$. 

We organize the work as follows: in Section \ref{sec:framework} we present the theoretical dark matter framework, introduce the BDM model in sub-section \ref{subsec:BDM} and compute the free streaming scale for different dark matter models in Section \ref{sec:lss}. In Section \ref{sec:lin_evol} we compute the CMB power spectrum using the perturbations of the BDM model and put constrains with Planck data, \ref{ssec:cmb_ps}, and show the cut-off scale induced by the BDM model in the matter power spectrum in sub-section \ref{ssec:mps}, and the mass function in Section \ref{ssec:preschechter} . We conclude in Section \ref{sec:conclusion}. Finally, we present the standard perturbations equations for BDM in appendix \ref{appendix:perturbation}. 
 
\section{General Dark Matter Framework}\label{sec:framework}

The DM particles are usually classified by their velocity dispersion given in terms of three broad categories: hot (HDM), warm (WDM) and cold (CDM) dark matter. The main difference between these three cases is the scale factor, $a_{nr}$, when the DM particles become nonrelativistic. In principle HDM is relativistic at all cosmological relevant scales, e.g. neutrinos, CDM has a small $a_{nr}$ (with $a_{nr}\ll a_{eq}$) while  WDM are particles in between. 
 
Relativistic particles with mass $m$  and a peculiar velocity, $v$, have a momentum ${p}=\gamma mv$, and energy $E^2=p^2+m^2$, where $\gamma\equiv 1/\sqrt{1-v^2}$. Solving for $v$ we get,
\begin{equation} \label{eq:vel_gral_eq}
 v  = \frac{p^2}{E^2} =\frac{p^2}{\sqrt{m^2 + p^2}}.
\end{equation}
For $p^2 \gg m^2$ the  velocity  is $v \sim 1$ and the particle is relativistic, while for $p^2\ll m^2$ we have $v \ll 1$ and the particle  is then non-relativistic.

In an expanding FRW Universe, the momentum of a relativistic particle redshift as  $p(a)= p_{\star} (a_\star/a)$, where $p_\star$ and $v_\star$, with the correspondent parameter $\gamma_\star \equiv (1 - v_\star^2)^{-1/2}$,  are a pivotal point condition for the momentum and the velocity at $a = a_\star$. Therefore, the velocity at all times in an expanding Universe evolves as
\begin{equation}\label{eq:veldm}
  v(a)  = \frac{ \gamma_\star v_\star(a_\star/a)}{\sqrt{1 + \gamma^2_\star v^2_\star(a_\star/a)^2}}.
\end{equation}
Eq.(\ref{eq:veldm}) describes the velocity evolution of a decoupled relativistic massive particle having a smooth transition from the relativistic limit $v\simeq 1$, for $a \ll a_\star$, to a non-relativistic behavior $v\simeq 0$, in the limit $a \gg a_\star$. This is a general evolution and it is valid for massive particles (WDM, CDM or massive HDM). 

 It is common to set  the epoch when the particle becomes non relativistic when  $p^2=m^2$, with $E^2=p^2+m^2=2m^2$, and from Eq.\eqref{eq:vel_gral_eq} the velocity is simply  $v(a_{nr})\equiv v_{nr}=1/\sqrt{2}$ with $\gamma_{nr}=\sqrt{2} $ and $\gamma_{nr} v_{nr}=1$. We have set the pivotal time at this epoch. i.e. $a_\star=a_{nr}, v_\star=v_{nr}$.  For a massive particle  that becomes non-relativistic at $a_{nr}$ the velocity at any time evolves as
\begin{equation}\label{eq:wdm_vel}
  v(a)  = \frac{ (a_{nr}/a)}{\sqrt{1 + (a_{nr}/a)^2}}.
\end{equation}
Notice here that for thermal particles the value of $a_{nr}$ can be directly related to the mass of the particle, a larger mass gives a smaller  $a_{nr}$ and become non-relativistic earlier.

In order to determine the evolution of  the energy density $\rho$, we take $\rho= \langle E\rangle\, n$ and the pressure   $P=\langle |\bar p|^2 \rangle n / 3\langle E \rangle$, with $n$ the particle number density, $\langle|\bar p|^2\rangle$  the average quadratic momentum and $\langle E\rangle$ the average energy of the particles. The equation of state (EoS)  $w\equiv \rho/P$  is given then by
\begin{equation}\label{eq:eos}
    \omega = \frac{ \langle |\bar p|^2\rangle}{3\langle E\rangle^2}= \frac{v(a)^2}{3}
\end{equation}
Integrating the continuity equation $\dot\rho=-3H(\rho+P)$ and   using Eq.\eqref{eq:veldm} and \eqref{eq:eos} we obtain the analytic evolution of  the background  $\rho_{\rm dm}(a)$
\begin{equation} \la{rm}
    \rho_{\rm dm}(a) = \rho_{\rm dm\star} \left( \frac{a}{a_\star} \right)^{-3} f(a)
\end{equation}
with 
\be\la{f}
f(a)= \le(\frac{a}{a_\star}\ri) ^{-1}\le(\frac{v_\star}{v}\ri) =  \gamma_\star^{-1}  \sqrt{1+\gamma_\star^2v_\star^2(a_\star/a)^2}.
\ee
Eqs.(\ref{rm}) and (\ref{f}) are valid for any value of $v_\star$ including the case of  a standard massive particle  with  $ v_\star=v_{nr}=1/\sqrt{2}$ at $a_\star=a_{nr}$.  From Eq.\eqref{f} we clearly see that $\rho_{\rm dm}(a_\star)=\rho_{\rm dm\star} $ since $f(a_\star)=\gamma_\star^{-1}\sqrt{1+\gamma_\star^2v_\star^2} =1$.  In the limit $a\ll a_\star$ we have  $\rho_{\rm dm}(a)=\rho_{\rm dm\star} (a/a_\star)^{-4} $, since $f\simeq (a_\star/a)$ and  $v_\star/ v \simeq 1$, showing that  $\rho_{\rm dm}(a)$ evolves as radiation, while in the late time limit $a\gg a_\star$ we have $f\simeq 1/\gamma_\star$ and  $\rho_{\rm dm}(a)=\rho_{\rm dm\star} (a/a_\star)^{-3} /\gamma_\star $.
Finally, in terms of present day values we obtain
\begin{equation} \la{rmo}
    \rho_{\rm dm}(a) = \rho_{\rm dm  o} \left( \frac{a}{a_o} \right)^{-3} \fr{f(a)}{f(a_o)}
\end{equation}
with $f(a)/f(a_o)\simeq 1$ for $a\gg a_\star$, thus $\rho_{\rm dmo}$ is the amount for dark matter density today.

\subsection{BDM Model}\label{subsec:BDM}
An interesting model not contained in the above description is our BDM model, previously introduced in \cite{delaMacorra:2009yb}. Here we just summarize the most important characteristics that will help us develop the present work. In the model of interest, the particles are relativistic for $a< a_c$ and they go through a phase transition at $a_c$, where the original elementary (massless or nearly massless) particles form bound states  which we call Bound Dark Matter BDM, similar as in QCD where quarks form  baryons and mesons. Clearly, the mass of the mesons and baryons, with masses of the order $\mathcal{O}$(GeV), do not correspond to the sum of the constituent quark masses, of which are of the order of $\mathcal{O}$(MeV).
 
\begin{figure}[t]
\centering
    \includegraphics[width=12cm]{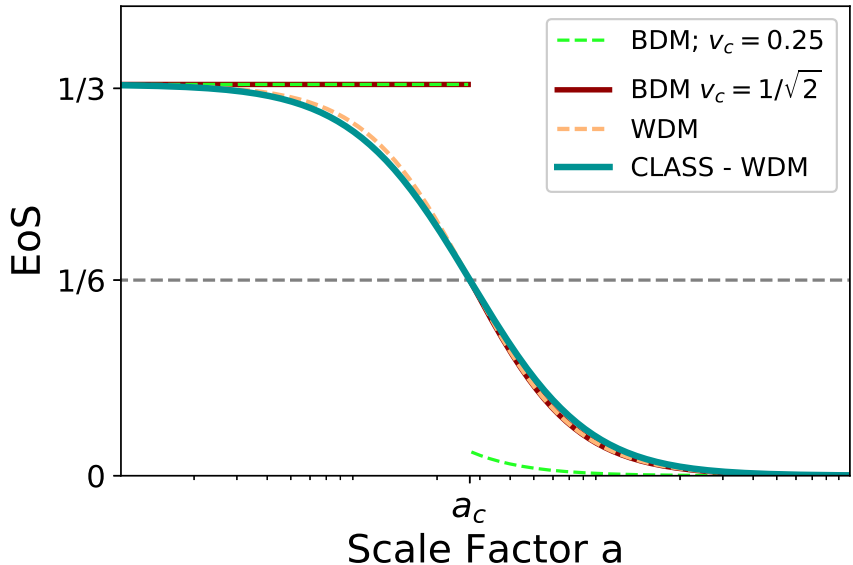}
    \caption{\footnotesize{ Plot of the equation of state for BDM and WDM. Continuous green and dashed yellow lines represent $\omega_{\rm bdm}$. The first one obtained from solving Boltzmann equations using CLASS, the second one is the analytic expression for the EoS of WDM (Eq.\eqref{eq:wdm_vel} and \eqref{eq:eos}) that became non-relativistic at $a_{nr} = a_c$. Red and gren dashed lines represent the BDM EoS. The red one has a initial velocity at $a_c$, $v(a_c) = 1/\sqrt{2}$. Notice that for $a > a_c$ the line goes over the yellow WDM line. The green line represent a BDM with a initial velocity $v_c = 0.25$, the evolution for $a < a_c$ overlap the other BDM model, but the EoS is $\omega_{\rm bdm}(a_c) = v_c^2/2$ and quickly evolve to zero after the transition.  }}
    \label{fig:eos} 
\end{figure}
 
We propose that  BDM particles  are relativistic  and massless  (or with a very small mass) for $a<a_c$ and  acquire a non-perturbative mass $m_{\rm bdm} \propto \Lambda_c$ at $a_c$,  due to the non perturbative effects of the underlying force with transition energy $\Lambda_c$.  Since this effect is a non-smooth transition, we expect the BDM particles to go from being masless, for $a<a_c$, to massive at $a_c$ (with a corresponding time $t_c$). Therefore, the velocity of the particle goes from  $v=1$, for $a<a_c$, to $v \rightarrow v(a_c) \equiv v_c$ (with $\gamma(a_c) \equiv \gamma_c$) and the evolution of its velocity and the EoS is  given by   
\begin{eqnarray}
  \begin{alignedat}{3}
   	\omega_{\rm bdm} & =  \frac{1}{3}, & \ v_{\rm bdm} &=1, & \quad  {\rm for} \, a < a_c  \\
  	\omega_{\rm bdm} &= \frac { v_{\rm bdm}^2} {3}, & v_{\rm bdm} &= \frac{ \gamma_{c} v_{c}(a_{c}/a)}{\sqrt{1 + \gamma^2_{c} v^2_{c}(a_{c}/a)^2}} & \quad {\rm for} \, a \geq a_c,   
   \end{alignedat}\label{eq:eos_bdm}   
\end{eqnarray}
where subindices $c$ denotes quantities evaluated at $a_c$. The case where $1/\sqrt{2} > v_c > 0$ describe a particle whose velocity has been suddenly suppressed due to bounding nature of the particle. The case where $1> v_c > 1/\sqrt{2} $, the particle acquired mass and the velocity is suppressed, but still being a relativistic particle, this particle become non-relativistic at $a_{nr} = \gamma_c v_c a_c $ and emulates WDM, and we do not consider  this case here. We plot $\omega_{\rm bdm}$ in Figure \ref{fig:eos}. The evolution of $\rho_{\rm bdm}$  before the phase transition at $a_c$ is that  of  relativistic energy density   while after the transition we have from Eq.(\ref{rm}) 
\begin{eqnarray}
  \begin{alignedat}{2}
	\rho_{\rm bdm}(a) & =  \rho_{\rm bdm  c} \left( \frac{a}{a_c} \right)^{-4}  &  \quad  {\rm for} \,\, a < a_c  \\
	\rho_{\rm bdm}(a) & = \rho_{\rm bdm  o} \left( \frac{a}{a_o} \right)^{-3} \fr{f(a)}{f(a_o)}  &  \quad {\rm for} \,\, a \geq a_c,   
   \end{alignedat}\label{eq:rho_bdm}   
\end{eqnarray}
with $ f(a)= \gamma_c^{-1} \sqrt{ 1+\gamma_c^2 v_c^2(a_c/a)^2}$, with $f(a_c)=1, f(v_c=0)=1$ and $f(a\gg a_c)=1/\gamma_c$. 

As seen in Eq.\eqref{eq:wdm_vel} a massive particle (WDM or CDM) becomes non-relativistic at $a_{nr}$ with $v_{nr}=1/\sqrt{2}$ and has only one free parameter, the scale factor $a_{nr}$, however in BDM the EoS has two free parameters, the time of the transition $a_c$ and the velocity dispersion $v_c$ at that time. We recover CDM if the transition happens at $a_c \ll 10^{-9}$, in this case, the velocity parameter is less important because the particle becomes non-relativistic in a very early stage of the Universe. HDM can also be described if the transition is of the order of $a_c \gtrsim \mathcal{O}(10^{-4}) $ and the particle is highly relativistic.

\section{Dark Matter models and Large Scale Structure} \label{sec:lss}

The distinctive feature of WDM or BDM particles is that both have non-negligible thermal velocities at early times that impact on large-scale structures. The relevance of the transition of the BDM particles, which is the novelty of this work, has very interesting astrophysical and cosmological consequences, on one hand, reduce the small scale structures. On the other hand, influence the internal structure of dark matter halos and on the galaxies they are hosting, which introduce a core in the density profile \cite{Mastache:2011cn, Maccio:2012qf}. The free-streaming washes out perturbations in the matter distribution below the $\lambda_{fs}$ scale \cite{Bond:1983hb, Benson:2012su} and leads to a very distinctive cutoff in the matter power spectrum, $P(k)$, at the corresponding scale.  Let us first determine the free streaming scale $\lambda_{fs}$ in WDM and BDM models and later we present a straight comparison between both BDM and WDM models.

The distinctive feature of WDM or BDM particles are non-negligible thermal velocities at early times.  Free streaming that washes out perturbations in the matter distribution below the free streaming scale  $\lambda_{fs}$ (\cite{Bond:1983hb, Benson:2012su} and leads to a very distinctive cutoff in the matter power spectrum, $P(k)$, at a corresponding scale and encoded in the valur of the half mode $k_{1/2}$
where the power spectrum is suppressed 59\% compared to a $\Lambda$CDM model. 

The BDM model is both, astrophysical and cosmological relevant. WDM, which impacts on both large-scale structure formation and the internal structure of dark matter haloes and the galaxies they are hosting, is the significant of the velocities of the BDM particles the novelty of this work. Let us first determine the free streaming scale $\lambda_{fs}$ in WDM and BDM models and later we will consider the power spectrum in both cases.

\subsection{Free-streaming scale}\label{ssec:freestreaming}

While DM particles are still relativistic, primordial density fluctuations are suppressed on scales of order the Hubble horizon at that time and it is given in term of  the free-streaming scale $\lambda_{fs}$. The free-streaming scale depends then on the time scale when the particles become  non-relativistic  given by   $a_{nr}$ in WDM  and  by  $a_c$ in our BDM model.  Free streaming particles have  an important impact since they suppresses the structure growth  and  impacts of in the CMB and Matter power spectrum. 
For standard thermal relics, the shape of the cutoff is therefore well characterized in the linear regime and there is an unambiguous relation between the mass of the thermal relic WDM particle and a well-defined free-streaming length. Note that we will also quote a freestreaming mass, which is the mass at the mean density enclosed in a half-wavelength given by the free streaming scale $\lambda_{fs}$.  We  study first the free-streaming of the WDM and BDM particles and later in Section(\ref{ssec:preschechter}) the structure abundances for masses using the Press-Schechter formalism.
The comoving  free streaming scale $\lambda_{fs}$ is defined by
\begin{equation}
  \lambda_{fs} = \int_0^{t_{eq}} \frac{v(t) dt}{a(t)} = \fr{2t_{c}}{a_{c}^2} \int_{0}^{a_{eq}} v(a) da, \label{eq:fs_general} 
\end{equation}
where we have assume a radiation dominated Universe  for $a\leq a_{eq}$ and  therefore  we have $t \propto a^2$. 
The free-streaming scale is defined by the mode $k_{fs}$  and the Jeans mass $M_{fs} $ contained in sphere of radius $\lambda_{fs}/2$ given by
\be
k_{fs}=\fr{2\pi}{\lambda_{fs}}, \;\;\;\;\;\;\;\;\; M_{fs} = \fr{4\pi}{3} \le(\fr{\lambda_{fs}}{2}\ri)^3 \rho_{mo}.
\ee
Haloes with masses below the free-streaming  mass scale will be suppressed.

\subsubsection{WDM scenario}

Let us now determine  the comoving  free streaming scale $\lambda_{fs}$, first for the fiducial CDM case. It is standard to separate the integral in the relativistic regime with  a constant speed $v=1$ for $a<a_{nr}$ and in the non-relativistic regime  $a>a_{nr}$ to take $v = a_{nr}/a$. With these choices of $v$  one gets the usual free streaming scale  
\bea
\nonumber
\lambda_{fs} (a_{eq})&=& \int_0^{t_{nr}} \frac{c dt}{a(t)} + \int_{t_{nr}}^{t_{eq}} \frac{v(t) dt}{a(t)}\\
 &=&\fr{2t_{nr}}{a_{nr}} \le[1+ \ln \le(\fr{a_{eq}}{a_{nr}}\ri) \ri]
\la{fs1}
\eea
where we used  that in radiation domination $t \propto a^2$ and   $t/t_{nr}=a^2/a^2_{nr}$. The first term in Eq.(\ref{fs1})
corresponds to the integration from $a=0$ to $a_{nr}$, while the second from  $a_{nr}$ to $a_{eq}$.
To compute the free-streaming for WDM is more accurate to us Eq.\eqref{eq:wdm_vel}  for the velocity, since it is valid for  all $a$. In this case we obtain a free streaming scale
 \bea
\lambda_{fs}(a) =\fr{2t_{nr}}{a_{nr}}\; \ln \le[\le(\fr{a}{a_{nr}}\ri)+\sqrt{\le(\fr{a}{a_{nr}}\ri)^2+1\,}\,\ri ] 
 \la{fs2a} 
 \eea
valid for all $a$.  Let us now evaluate Eq.(\ref{fs2a})  at $a_{eq}$  and  take the limit $a_{nr}/a_{eq} \ll 1$ to get
\be
\lambda_{fs}(a_{eq})\simeq  \fr{2t_{nr}}{a_{nr}}  \le(\ln[2]  +  \ln \le[\fr{a_{eq}}{a_{nr}}\ri] \ri) = \fr{2t_{nr}}{a_{nr}} \ln \le[\fr{2\, a_{eq}}{a_{nr}}\ri] 
\la{fs4}\ee
with $\ln[2]\simeq 0.69$.  Eq.(\ref{fs2a}) or its limit Eq.(\ref{fs4}) should be used instead of Eq.(\ref{fs1}) since they capture the full evolution of the velocity $v(a)$ of a massive particle  given in Eq.(\ref{eq:wdm_vel}).  We can easily estimate eq.(\ref{fs4}) by assuming an universe radiation dominated for $a\leq a_{eq}$ and matter dominated  for $a\geq a_{eq}$.  In such a case we can express $2t_{nr}=1/H_{nr} $ with  $H_{nr}=  (a_o/a_{nr})^2(a_{eq}/a_o)^{3/2} H_o$ and  $c/H_o=2998 h^{-1}$Mpc  the Hubble length  to  obtain
\be
\fr{2t_{nr}}{a_{nr}}=11.14  \le(\frac{3411+1}{z_{eq}+1} \ri) \le(\frac{3.18\times10^{-8}}{a_{nr}}\ri) \; {\rm h^{-1} kpc}
\la{fs4b}\ee
The free-streaming, $\lambda_{fs}$, and the Jeans mass take the following values,
\begin{eqnarray}\la{fs5}
    \lambda_{fs}^{\rm wdm} &=& 0.11  \left(\ln\left[\le( \frac{3411+1}{z_{eq}+1}\ri)\le(\frac{3.18\times10^{-8}}{a_{nr}}\right)\ri] + 9.82 \right) \frac{\rm Mpc}{\rm h}   \\
    M_{fs}^{\rm wdm} &=& 1.97\times10^7\le( \frac{\Omega_{dm0}h^2}{0.12}\ri)\left( \frac{\lambda_{fs}}{1.1}\frac{\rm h}{\rm Mpc} \right)^3 M_\odot .
    \la{fs6}
\end{eqnarray}

\subsubsection{BDM scenario}

Let us now determine the free streaming scale for our BDM model.  The  velocity of the particle is given by Eq.(\ref{eq:eos_bdm}), which takes into account the transition for BDM, this leads to the free streaming scale
\be
\lambda_{fs}^{\rm bdm} (a_{eq})= \int_0^{t_c} \frac{c dt}{a(t)} + \int_{t_c}^{t_{eq}} \frac{v(t) dt}{a(t)}
\ee
giving 
\begin{eqnarray}
     \lambda_{fs}^{\rm bdm} (a_{eq})&=& \fr{2t_{c}}{a_{c}^2} \left[ \int_0^{a_c} da + \int_{a_c}^{a_{eq}} v(a) da  \right] \label{eq:fs_bdm0} \\
     &=& \fr{2t_{c}}{a_{c}} \le( 1 +  \gamma_c v_c \textrm{ln} \le[\fr{ 1+ \sqrt{1 + \gamma_c^2 v_c^2(a_c/a_{eq})^2  }}{ {(1 + \gamma_c)(a_c/a_{eq})} } \ri] \ri)  \la{fs2}  \nonumber \\
     &\simeq& \fr{2t_{c}}{a_{c}} \le( 1 + v_c \gamma_c  \textrm{ln} \le[ \frac{2}{(1 + \gamma_c)}\frac{a_{eq}}{a_c} \ri]  \ri), \la{fs3}
\end{eqnarray}
where $t_c$ is the time corresponding to the transition $a_c$. In the last equation we assume that $a_c \ll a_{eq}$ and  $2t_{c}=1/H_{c} $, $H_{c}=  (a_o/a_{c})^2(a_{eq}/a_o)^{3/2} H_o$  and $c/H_o=2998 h^{-1}$Mpc.  Clearly the value of $v_c$ in eq.(\ref{fs3}) has a huge impact on the resulting free streaming scale    and taking the limit $v_c=0$  eq.(\ref{fs3}) becomes  equivalente to eq.(\ref{fs4b}) and subsituting $a_{nr}, t_{nr}$ with  $a_{c}, t_{c}$, i.e. 
\be
    \lambda_{fs}^{\rm bdm} \simeq \fr{2t_{c}}{a_{c}} = 11.14  \le(\frac{3411+1}{z_{eq}+1} \ri) \le(\frac{3.18\times10^{-8}}{a_{c}}\ri) \; {\rm h^{-1} kpc}.
\la{lc}
\ee
If we compare eq.(\ref{lc}) with eq.(\ref{fs5}) we see that the effect of having $v_c=0$ reduces the free streaming scale by a factor close to $10$ and the corresponding Jeans Mass by a factor of $10^3$.

For example,  if we take a BDM  that becomes no-relativistic at the same scale factor as  a 3 keV mass WDM we find for $v_c=0$ that $\lambda_{fs} = 0.01$ Mpc/h, with a contained mass of $M_{fs} =(4\pi  \rho_{dm} /3)(\lambda_{fs}/2)^3 = 2.08 \times 10^{4} \ M_\odot/ h^3$, in contrast with $\lambda_{fs} = 0.1$ Mpc/h and  $M_{fs} = 1.97 \times 10^{7} \ M_\odot/h^3$ for WDM.  
Clearly the impact of $v_c$ can be huge  in structure formation  by
be severely reducing  (three orders of magnitude in our previous example )  that amount of mass contained on a structure of radius $r\sim \lambda_{fr}/2$   in BDM compared to a WDM, while having the same  scale factor  $a_{nr}=a_c$ when they become nonrelativistic.

\section{Comparing WDM  vs BDM}

In this section we compare BDM and WDM models and determine the scale factor when they particles become non relativistic.  As the Universe expands the temperature redshifts as $T\propto 1/a$  with velocity $v(a)$ given by eq.\eqref{eq:wdm_vel} and eventually the WDM particle  becomes  non-relativistic when $p^2 = m^2$ given by $a_\textrm{nr}$ with  a velocity dispersion  $v(a_{nr})=1/\sqrt{2}$.  In BDM the  transition from relativistic particles to non-relativistic  takes place at $a_c$ with a velocity dispersion $v_c(a_{c}) $ which is  not well defined
since  the transition is due to the underlying non-perturbative mechanism that generates the phase transition and is therefore  not well determined with a  range of values of $1/2 \geq v_c^2\geq 0$. 
For example the masses of the baryons and mesons in of the Standard Model (SM) of particles are due to the binding energy  of the  the strong (QCD) force generated once the QCD force becomes strong with masses of the order of $m \sim  O(\Lambda_{QCD})\simeq 200 MeV$. It is worth keeping in mind, however, that our  BDM is by hypothesis not contained in the SM. 

 In order to relate the mass of the WDM particle  to  the scale factor $a_{nr}$  let us take the non-relativist limit  of the EoS $w$  and we approximate  $w=p/\rho=T/M=v^2/3$,  with  $T$ the temperature, $M$ its mass and $v(a)$ its velocity, and we used $T(a)=M v(a)^2/3$ and
  valid for $a\geq a_{nr}$. At $a=a_{nr}$ we have  $ v(a_{nr})=1/\sqrt{2}$, $T(a_{nr})=M/6$ and $w(a_{nr})=1/6$. 
  
 The relativistic energy density is then given by $\rho =(\pi^2 g_x/30)  T^4$ for  $a\leq a_\textrm{nr}$  with $\rho_{dm} (a_{nr}) = (\pi^2 g_x/30)  (M v(a_{nr})^2/3)^4=  (\pi^2 g_x/30)  (M v^2(a_{nr})/3)^4$  at  $ a_\textrm{nr}$ while the evolution of $\rho_{dm}(a)$ for $a\geq a_\textrm{nr}$ is given in Eq.(\ref{rmo}).   
We equate the energy densities of these two region at $a_{nr}$, i.e.  $\rho(a_{nr})=(\pi^2 g_x/30)  (M/6)^4
=\rho^{\rm wdm}_o(a_o/a_{nr})^3 f(a_{nr})/(fa_o)$  to obtain
\be
\fr{a_{nr}}{a_o} =  5.6  \times 10^{-8}   \le(\fr{\Omega_{dmo}h^2}{0.120}\ri)^{1/3} \le(\fr{3\,keV}{M}\ri)^{4/3} \le(\fr{7/4}{g}\ri)^{1/3}   . \\
\la{eq:anrao}
\ee
We can now relate the time when two  different WDM models become non-relativistic  with the same amount of WDM today
and we obtain 
\be
\fr{a_{nr}}{a_{nr}'} =\le( \fr{T'}{T}\ri)^{4/3}=\le( \fr{M'}{M}\ri)^{4/3}  \le( \fr{v(a_{nr})}{v'(a'_{nr})}\ri)^{8/3} 
\la{mm}\ee
where the last equality allows for a the possibility of having a different value of the velocity dispersion $v$ at $a_{nr}$, relevant for BDM models. 
If  both DM models have  $v(a_{nr})=v'(a'_{nr})=1/\sqrt{2} $ (i.e. they are two thermal WDM models)  Eq.(\ref{mm})  reduces to 
\be
\fr{a_{nr}}{a_{nr}'} =\le( \fr{M'}{M}\ri)^{4/3},
\la{mm2}\ee
and we can estimate the value of $a_{nr}$ given  $a_{nr}'$ in term of the masses of the WDM particles. On the other hand,  Eq.(\ref{mm}) also shows that two DM models with same mass $M=M'$ but with different  values of $v(a_{nr})$ give
\be
\fr{a_{nr}}{a_{nr}'} =\le( \fr{v(a_{nr})}{v'(a'_{nr})}\ri)^{8/3}. 
\la{mm3}\ee
One of these two models may be WDM and the other a BDM which has a reduced $v_{nr}$  at $a_{nr}$ (i.e. $v_c(a_c)$ for BDM ).

Let us now compare $\lambda_{fs}$ in WDM and BDM models given in Eq.\eqref{fs4} and Eq.\eqref{fs3} both  with the same $a_{eq}$ and 
using $t_{nr}=(a_{nr}^2/a_{eq}^2) t_{eq}, \;t_c=(a_c^2/a_{eq}^2) t_{eq}$ and we get
\be
\fr{ \lambda_{fs}^{wdm}}{ \lambda_{fs}^{bdm}}=\fr{a_{nr}}{ a_c} \fr{ [ 0.69+\; \ln(a_{eq}/a_{nr}) ]}{\le( 1 + v_c \gamma_c  \ln \le[ \frac{2}{(1 + \gamma_c)}\frac{a_{eq}}{a_c} \ri]\ri)  }. 
\la{fswfsb}\ee
%
Taking the limiting case $v_c=0$ and the condition $ \lambda_{fs}^{wdm}=  \lambda_{fs}^{bdm}$  gives  
  \be
a_c=a_{nr}[0.69+\; \ln( a_{eq}/a_{nr})]. 
 \la{fswfsb2}\ee
Notice that $a_c$ is an order of magnitude larger than $a_{nr}$. 
Therefore a  BDM particle that becomes non relativistic at $a_{nr}= 3.12\times10^{-7} $  and has $v_c=0$  has an equivalent mass   as WDM particle with $a_{nr}= 3.18\times10^{-8} $  and using eq.(\ref{mm2}) we  obtain a mass  
\be
M_{\rm bdm}= 532.97\, \rm{eV}\; \le( \fr{M }{3 {\rm keV}} \ri) \le(\fr{a_{nr}}{ 3.18 \times 10^{-8}}\ri)^{-3/4} \le(\fr{3.12\times10^{-7}}{a_c}\ri)^{-3/4},
\ee
i.e. the BDM particle with  mass of $532.97\, \rm{eV}$ has the same free streaming scale as a WDM with a mass of 3keV, so these two models suppress the growth of the same halo structures.

To conclude, a BDM particle that becomes non-relativistic at a scale of $a_c/a_o=3.12\times10^{-7}$, with $v_c=0$  has  the same free streaming scale $\lambda_{fs} $ and suppression of  halo mass $M_{fs}$   as a WDM particle with a mass $m_{wdm} = 3 \, keV$  while the WDM  becomes non-relativistic much earlier, at  $a_{nr}/a_o=3.18 \times 10^{-8} $. 
While at large scales $a\gg a_c$ ($a\gg a_{nr}$) for BDM (WDM), the structure formation is the same as in CDM, scales below the free-streaming scale are suppressed and modulated by the velocity dispersion $v_c$ ($v_{nr}$).

  \begin{figure*}
  \centering
  \begin{minipage}[t]{.8\textwidth}
       \includegraphics[width=\textwidth]{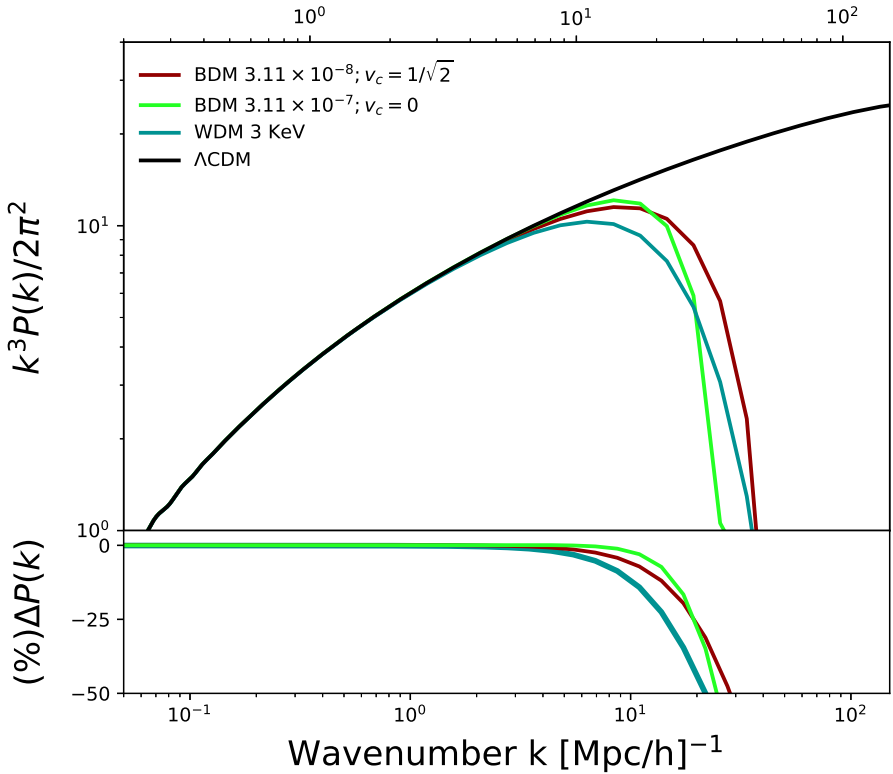}
     \caption{\footnotesize{ Matter Power Spectrum comparing WDM with BDM. The blue line corresponds to 3 keV WDM with $a_{nr} = 3.18\times10^{-8}$ obtained from CLASS. We compare with different BDM choices: Red line: For $a_c = a_{nr}$ and $v_c = 1/\sqrt{2}$. Green line: For a latter transition but $v_c = 0$ in order to preserve same free-streaming scale as WDM. }}\label{fig:WDM_as_BDM}
  \end{minipage} \hfill
\end{figure*}

\begin{table}
    \centering
\begin{tabularx}{\linewidth}{l|ccccc}
                                                    & $a_{nr} \, (a_c)$ & $k_{1/2}$ [h/Mpc]  & $\lambda_{fs}$ [Mpc/h]  & $k_{fs}$ [h/Mpc]  & $M_{fs}$ $[M_\odot / h^3]$ \\
\hline \hline
WDM 3keV                                 & $3.18\times 10^{-8}$ & 21.86   & 0.104 & 60.4   & $1.97\times 10^7$ \\
${\rm BDM}_1$ $v_c = 1/\sqrt{2}$              & $3.18\times 10^{-8}$ & 27.52   & 0.105 & 59.6   & $2.04\times 10^7$ \\
${\rm BDM}_2$  $v_c = 0$                          & $3.18\times 10^{-7}$ & 24.83   & 0.104 & 60.2   & $2.08\times 10^7$ \\
${\rm BDM}_3$  $v_c = 0$                          & $3.18\times 10^{-8}$ & 218.6 & 0.011 & 592.9 & $2.08\times 10^4$ \\
${\rm BDM}_4$  $v_c = 0$                          & $2.66\times 10^{-6}$ & 2.75 & 1.114 & 5.6 & $2.41\times 10^{10}$ \\

\end{tabularx}
\caption{In this table we show the comoving free streaming scale, $\lambda_{fs}$ (Eq\eqref{fs4} and Eq.\eqref{fs3} for WDM and BDM, respectively), the correspondent mode $k_{fs}$ and Jeans mass, $M_{fs}$ (Eq.\eqref{kMfs}) for different models, we also show the $k_{1/2}$ obtained from numerical simulations using CLASS. WDM correspond to a $m_{\rm wdm} = 3$ keV. BDM$_1$ and BDM$_2$ demonstrate that the $\lambda_{fs}^{\rm wdm}$ could be obtained with a BDM model with two $a_c$ and $v_c$. BDM$_3$ show the strong influence of $v_c$ on the Jeans mass, and BDM$_4$ is the limiting case according to the Montecarlo results. }
\label{tab:results}
\end{table}

\begin{figure}[t]
  \centering
    \includegraphics[width=10cm]{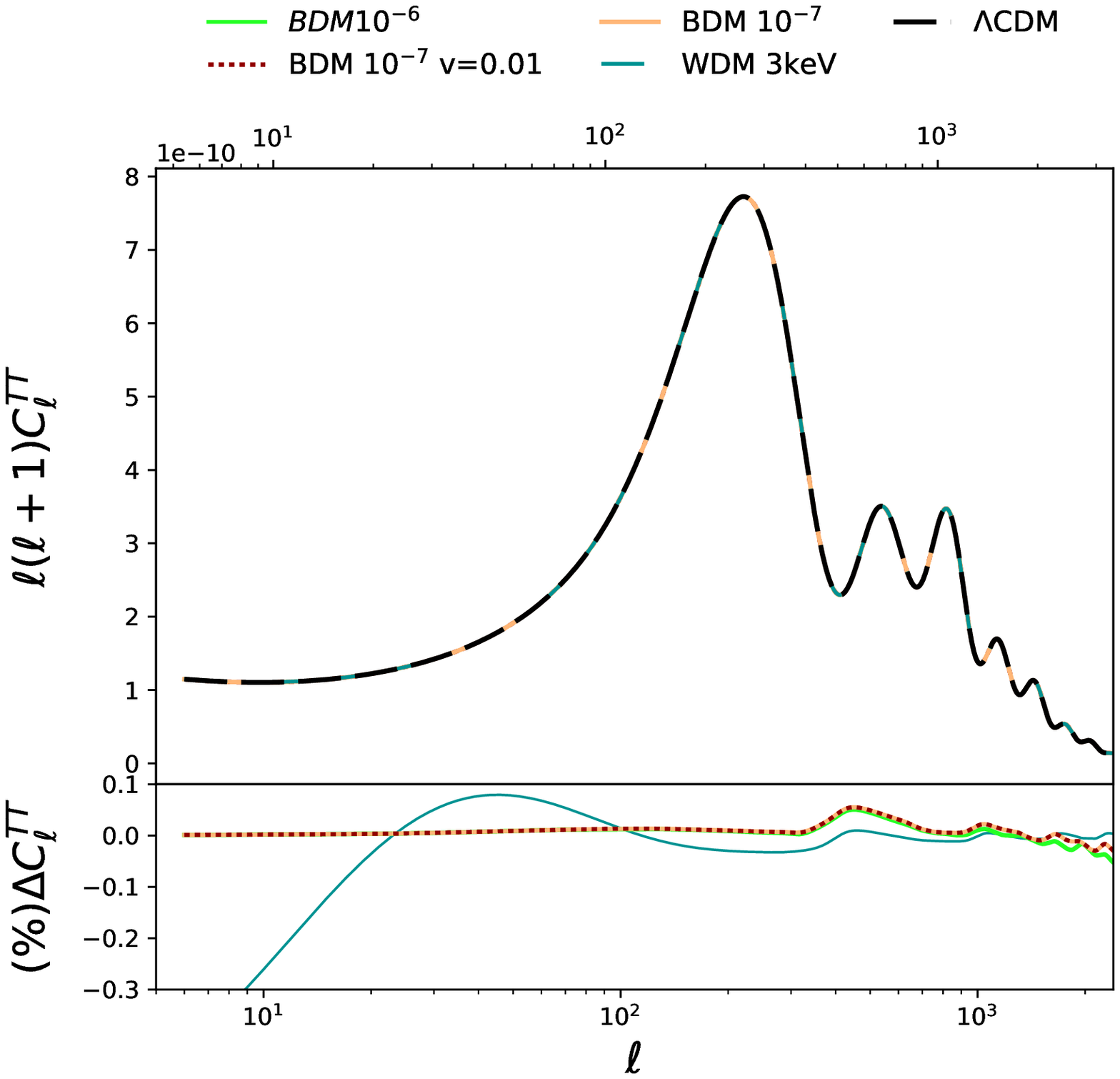}
    \caption{\footnotesize{
        {\bf Top panel}. We plot the CMB power spectrum for CDM (black solid line), 3 keV mass WDM (blue dashed line) and the BDM models with a different combination of parameters, we plot a BDM model with \{$a_c = 10^{-6}, v_c = 1/\sqrt{2} $ \} (green line), and $a_c = 10^{-7}$ for different values for the velocity $v_c = 1/\sqrt{2}$ (orange) and $v_c \sim 0$ (dotted red). Notice that changing the velocity parameter, $v_c$, does not significantly change the power spectrum. {\bf Bottom panel}. Percentage difference between CDM with WDM and BDM models.}}  \label{fig:power_spec}
 \end{figure}


\section{Linear Perturbation Theory}\label{sec:lin_evol}

We have computed the free-streaming, which is the mechanism that erases fluctuation below the $\lambda_{fs}$ scale and defines the minimum mass object that can collapse and form in the Universe through the Jeans mass, $M_{fs}$. Now it is interesting to compute the density perturbations in the BDM scenario to have a better sense on the cut-off scale. First, we compute the CMB power spectrum in the fluid approximation and run Montecarlo chains to constrain the $a_c-v_c$ parameter space \ref{ssec:cmb_ps}. Then, we compute the linear matter power spectrum and show the cutoff scale as a function of the time of the transition $a_c$ and the thermal speed at the transition, $v_c$ \ref{ssec:mps}.  Throughout this paper, we adopt Planck 2018 cosmological parameters \cite{Aghanim:2018eyx}. For the several simulations we adopt a flat Universe with $\Omega_c h^2 = 0.12$, and $\Omega_b h^2 = 0.02237$ as the CDM matter and baryonic omega parameter. $h = 0.6736$ is the Hubble constant in units of 100 km/s/Mpc, $n_s = 0.965$ is the tilt of the primordial power spectrum. $z_{\rm reio} = 7.67$ is the redshift of reonization and $\ln(10^{10} A_s) = 3.044$, where $A_s$ is the amplitude of primordial fluctuations.

\subsection{CDM Power Spectrum with Planck data}\label{ssec:cmb_ps}

In Figure \ref{fig:power_spec} we show the CMB power spectrum obtained with the CLASS code \cite{Blas:2011rf} taking into account the BDM perturbations, see Appendix \ref{appendix:perturbation}. We show the power spectrum for two different values of $a_c = \{10^{-6}, 10^{-7}\}$ and $v_c = \{0.01, 1/\sqrt{2}\}$. The smaller the value for $a_c$ and the smaller the velocity $v_c$ implies that BDM is more like CDM, therefore the difference between the curves is less notorious. We also compare the curves with a WDM particle with a mass of 3 keV. We notice that the effect of changing the initial velocity, $v_c$, is negligible for the CMB power spectrum. The percentage difference between $\Lambda$-BDM and $\Lambda$-CDM power spectrum, shown in the bottom panel of Figure\ref{fig:power_spec}  is less than 0.1\% for the BDM cases. The CMB power spectrum barely increased the height of the acoustic peaks, mainly because of the increase in the free-streaming smooth out perturbations and increase the acoustic oscillations.  

\begin{figure*}[t!]
\centering
  \begin{minipage}[t]{.85\textwidth}
     \includegraphics[width=\textwidth]{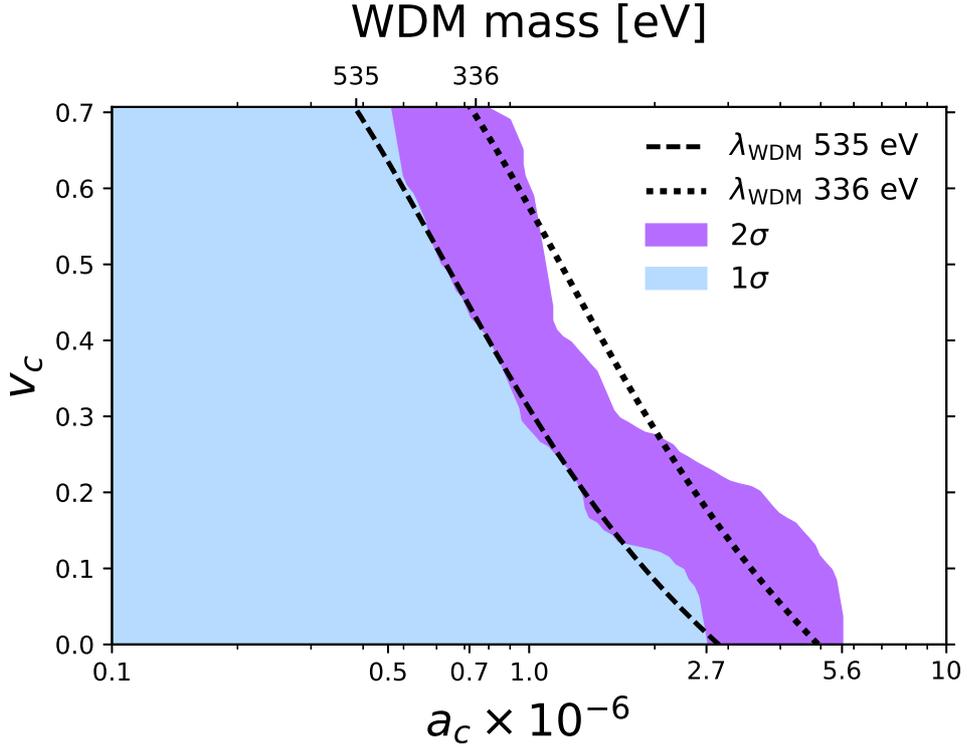}
    \caption{\footnotesize{ Likelihoods for BDM parameters, $a_c$ and $v_c$, from MonteCarlo simulations using MontePython with Plank 2018, type Ia SN, and BAO data. Light blue region represent $1\sigma$ likelihood, where largest values of $a_c$ are still valid. Larger value of $v_c > 1/\sqrt{2}$ are no relevant for BDM because the particle is relativistic after the transition. Purple region is the $2\sigma$ likelihood. Lines represent the different values of $a_c$ and $v_c$ that preserve the free-streaming of different WDM mases: dashed line for 535.4 eV mass, dotted line for 336.3 keV. This two last cases represent the lower bounds for WDM obtained by fitting to the border of the $1\sigma$ and $2\sigma$ of the respective likelihoods.  }}  \label{fig:likelihood}
  \end{minipage}
\end{figure*}

We perform the analysis by constraining our model using the 2018 Planck CMB likelihoods \cite{Aghanim:2018eyx}, we also include tBAO measurements\cite{Anderson:2013zyy}, and the JLA SNe Ia catalog\cite{Betoule:2014frx} to provide a reasonable representation of the degeneracies. In Figure \ref{fig:likelihood} we show the likelihood for the Montecarlo run using Montepython. The shadow areas correspond to $1\sigma$ and $2\sigma$ likelihoods, this is, we have no central values for the parameters, instead we are only able to put lower bounds to $a_c$. Smaller values of $a_c < 10^{-7}$ are within $1\sigma$ likelihood, and this is reasonable because smaller values of $a_c$ implies early transitions and bigger masses for the dark matter particle, just as CDM. Figure \ref{fig:likelihood} shows the smaller bounds in the $a_c - v_c$ parameter space, nevertheless show an interesting connection with WDM. We can compute $a_{nr}$ for a given mass for the WDM particle using Eq.\eqref{mm}, and Eq.\eqref{fswfsb} is the relation between $a_c$ and $v_c$ for a given $a_{nr}$, therefore the different lines in Figure \ref{fig:likelihood} represent the different values of $a_c$ and $v_c$ that preserve the free-streaming scale of a specific WDM particle with a non-relativistic transition at $a_{nr}$. Therefore the boundary of the $1\sigma$ ($2\sigma$) likelihood for the $a_c - v_c$ fits better to $a_{nr} = 3.16 \times10^{-7} \, (5.87\times10^{-7})$ that corresponds to a WDM mass of $m_{\rm wdm} = 535$ eV (336 eV).

The lower constrains for $a_c$ that corresponds to a $v_c = 0$ for the BDM model given the 1$\sigma$ and 2$\sigma$ likelihoods are
\begin{equation}
\begin{aligned}
   a_c \leq 4.59 \times 10^{-7}  \quad \quad {\rm for \;}& v_c = 1/\sqrt{2}  \quad\quad  &{\rm 1 \sigma \; likelihood}   \\
   a_c \leq 2.66 \times 10^{-6}  \quad \quad {\rm for \;}& v_c = 0  \quad\quad              &{\rm 1 \sigma \; likelihood}   \\
   a_c \leq 7.19 \times 10^{-7}  \quad \quad {\rm for \;}& v_c = 1/\sqrt{2}   \quad\quad &{\rm 2 \sigma\; likelihhod}   \\
   a_c \leq 5.62 \times 10^{-6}  \quad \quad {\rm for \;}& v_c = 0  \quad\quad               &{\rm 2 \sigma\; likelihhod}  
\end{aligned}
\end{equation}

\subsection{Matter Power Spectrum: WDM} \label{ssec:mps}
 
\begin{figure*}
\centering
    \begin{minipage}[t]{.49\textwidth}
     \includegraphics[width=\textwidth]{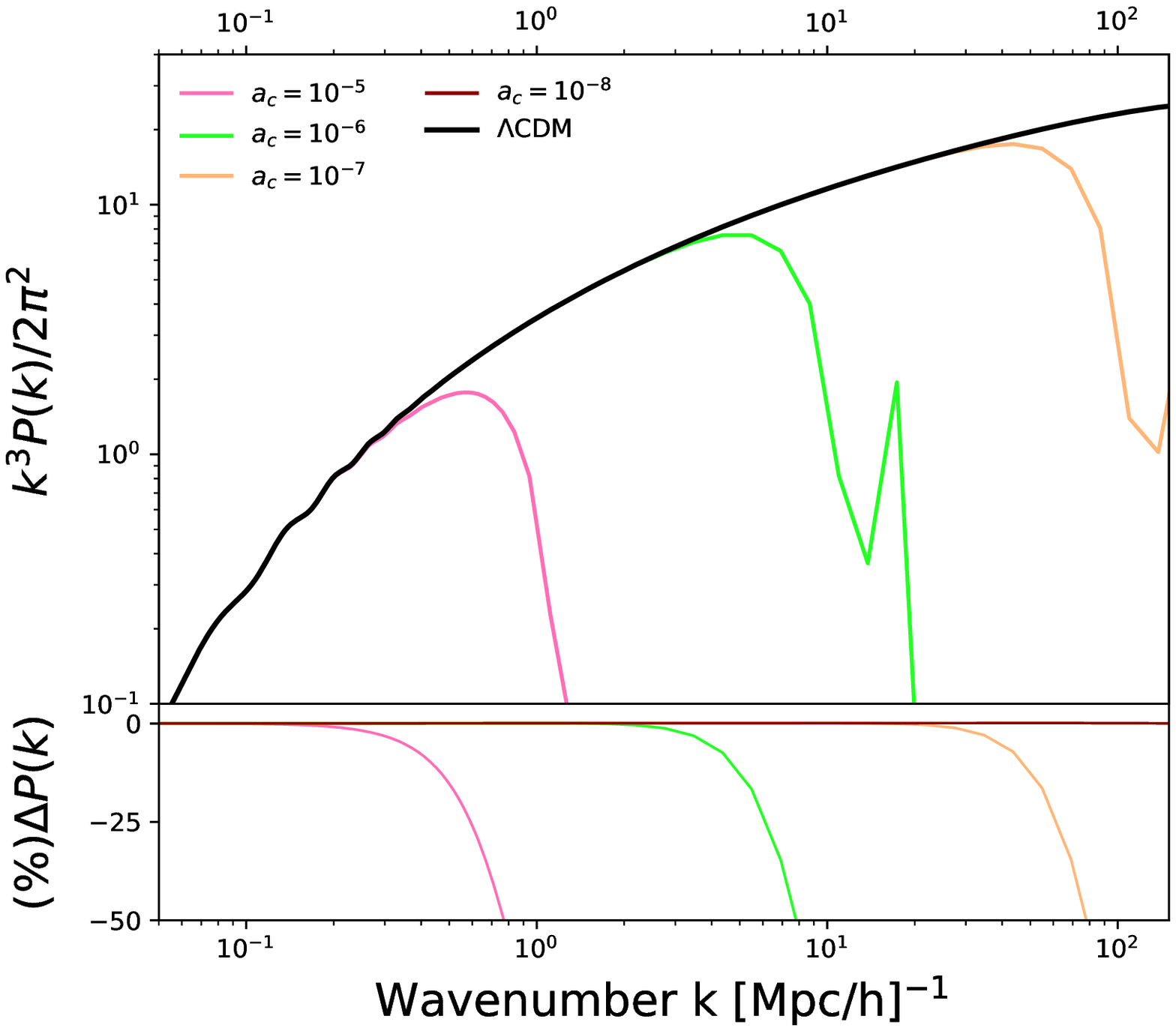}
    \caption{\footnotesize{ {\bf Top panel}. Plots of linear dimensionless matter power spectra for the CDM (black solid line) and BDM models for different values of $a_c$ and fixed $v_c$. The smaller $a_c$ the closer to CDM results. The velocity for all this cases is a very small numerical value close to zero, $v_c \sim 0$. {\bf Bottom panel}. We show the percentage difference, $\Delta P(k)$, between CDM and the models mentioned above. Notice that $k_{1/2}$ is defined when the difference between different matter power spectrum reaches 50\% difference.  }}  \label{fig:mps_fix_vc}
  \end{minipage}
  \hfill
  \begin{minipage}[t]{.49\textwidth}
      \includegraphics[width=1\textwidth]{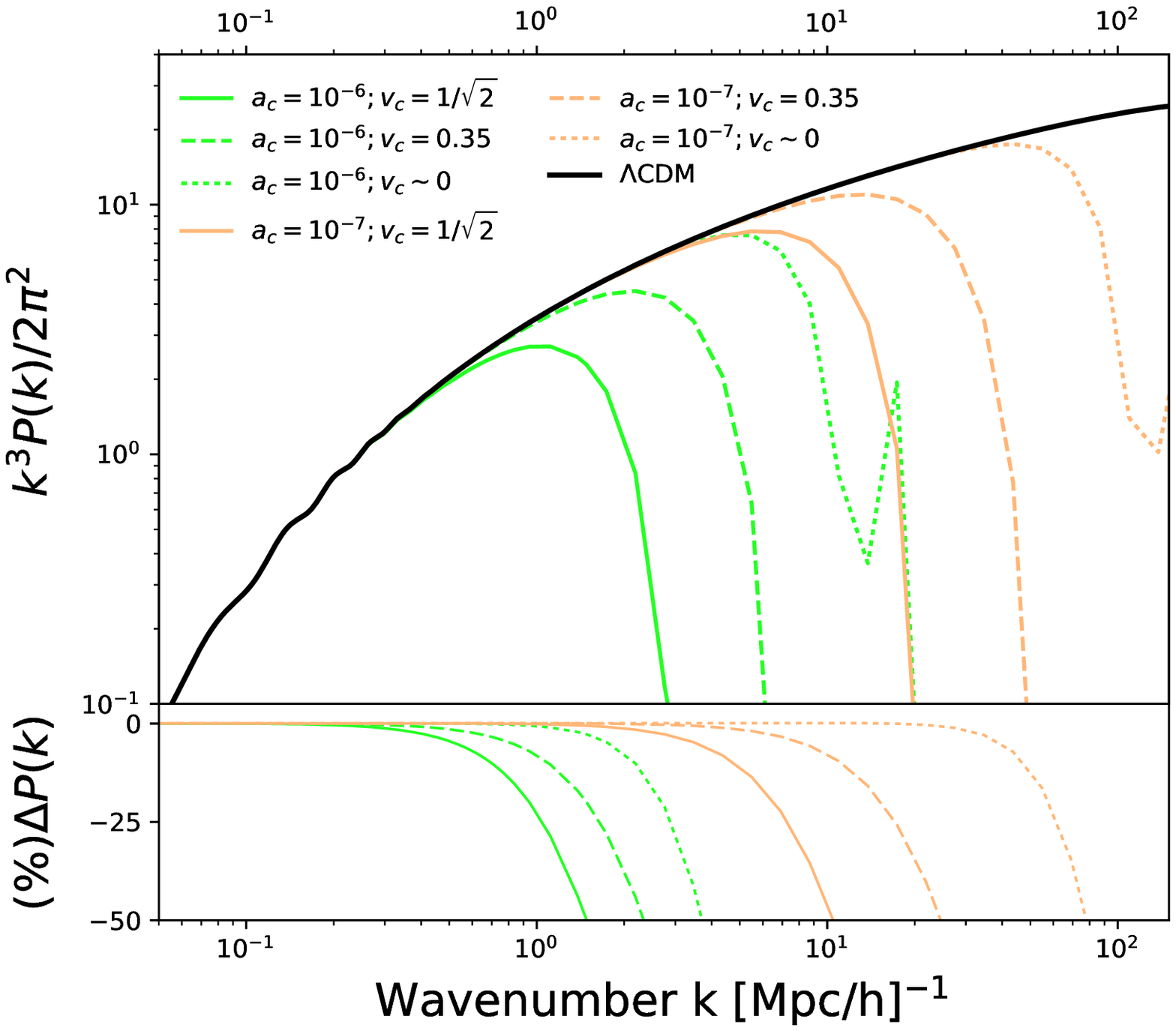}
     \caption{\footnotesize{ {\bf Top panel}. Plots of linear dimensionless matter power spectra for the CDM (black solid line) and BDM models using different values of $v_c = \{1/\sqrt{2}, 0.35, 0\}$ and $a_c = \{10^{-6}, 10^{-7}\}$. Clearly the value of $v_c$ affects the cut-off in the matter power spectrum, the smaller the value of $v_c$ the larger the cut-off scale, since the DM became non-relativistic faster. {\bf Bottom panel}. We show the percentage difference, $\Delta P(k)$, between CDM and the above mentioned models. Notice that  $k_{1/2}$ is define when the difference between different matter power spectrum reaches 50\%difference. } }\label{fig:mps_fix_ac}
  \end{minipage}
\end{figure*}

We now compute the CMB and  matter power spectrums  for  WDM and BDM scenarios  using the code CLASS ( see Appendix \ref{appendix:perturbation} for more details).  
The parametrization of the MPS along with the cut-off scale can be found for WDM \cite{Viel:2005qj, Bode:2000gq}. 

\begin{equation}\label{eq:fitted_viel}
  T _ { \mathrm { X} } ( k ) = \left[ \frac { P _ { \mathrm { lin } } ^ { \mathrm { X } } } { P _ { \mathrm { lin } } ^ { \mathrm { \Lambda CDM } } } \right] ^ { 1 / 2 } = \left[ 1 + ( \alpha k ) ^ { 2 \mu } \right] ^ { - 5 / \mu }
\end{equation}
with $\mu = 1.12$ and 
\be
\alpha= 0.049 \le( \fr{m_x}{1 keV} \ri)^{1.11}   \le( \fr{\Omega_x}{0.25} \ri)^{0.11}  \le( \fr{h}{0.7} \ri)^{1.22}  h^{-1} Mpc
\ee
or
\be
\alpha= 0.024 \le( \fr{m_x/T_x}{1 keV /T_\nu} \ri)^{-0.83} \le( \fr{w_x}{0.25(0.7)^2} \ri)^{-0.16}Mpc
\ee \label{alpha2}
with $X$ being the dark matter particle, either WDM or BDM.  

The same parametrization can be use for BDM particles, this  is define
We found that this parametrization is valid for$k \leq k_{1/2}$, where $k_{1/2}$  is obtained by setting $T_X(k)^2 =1/2$, we therefore have
\begin{equation} \label{eq:k_half}
    k _ { 1 / 2 } = \frac { 1 } { \alpha } \left[ \left( \frac { 1 } { \sqrt { 2 } } \right) ^ { -\mu / 5 } - 1 \right] ^ { 1 / 2\mu },
\end{equation}
for smaller scales the difference between numerical MPS and the parametrization of Eq.\eqref{eq:fitted_viel} became bigger, but less than 50\%, mainly because the cut-off of the BDM model are stepper than the ones obtained from WDM.

\subsection{Matter Power Spectrum: BDM}

In Fig \ref{fig:mps_fix_vc} and \ref{fig:mps_fix_ac}  we plot the linear dimensionaless power spectrum. The effect of the free-streaming (computed in subsection \ref{ssec:freestreaming}) for BDM particles is to suppress structure formation below a threshold scale, therefore the matter power spectrum show a cutoff at small scales depending the value of $a_{c}$ and $v_{c}$. From Figure \ref{fig:mps_fix_vc} one can notice that smaller the scale of the transition, $a_c$, the power is damped at smaller scales, for transitions at $a_c \lesssim 10^{-9}$ BDM model is indistinguishable from CDM at observable scales, $k \sim \mathcal{O} (100)$ Mpc/h. The scale of the transition would correspond to a WDM mass of $m_{\rm wdm} \sim 40$ keV.

The novelty in this work comes with the relevance that the velocity $v_c$ takes for LSS, in Figure\ref{fig:mps_fix_ac} we show that smaller values of $v_c$ implies a cooler dark matter, therefore a cutoff at smaller scales, and for a  transition  at fixed $a_c$ the cutoff scale in the matter power spectrum can vary an order of magnitude.  For instance, we can have the same free-streaming scale for two different massive particles, for  example a particle having a transition at $a_c = 10^{-8}, v_c=1/\sqrt{2}$ has a similar  free-streaming scale as a particle  with  $(a_c = 10^{-7}, v_c \sim 0)$.

The parameterization of the MPS along with the cutoff scale can be found for WDM \cite{Viel:2005qj}. The same parameterization can be used for BDM particles. The cut-off of the power spectrum depends on the parameter $\alpha$, for the BDM case its value depends on $a_c$ and $v_c$, and it can be computed using the relationship $T/m=v^2/3=v_c^2 (a_c/a)^2$ valid after the phase transition at $a_c$  and $T_\nu=(a_o/a)T_{\nu\,o}$ in Eq.(\ref{alpha2}) and we obtain

\be\la{alpha5}
\alpha= 0.024 \le( \fr{a_c v_c^2}{3}  \fr{1 keV } {T_{\nu o}} \ri)^{0.83} \le( \fr{w_x}{0.25(0.7)^2} \ri)^{-0.16}Mpc.
\ee
Notice that for the limiting case $v_c=0$ we have $\alpha=0$ and $T_{X}=1$ implying that  the power spectrum at $a_c$ has a steep decrease.

\begin{figure*}
\centering
    \begin{minipage}[t]{.8\textwidth}
     \includegraphics[width=\textwidth]{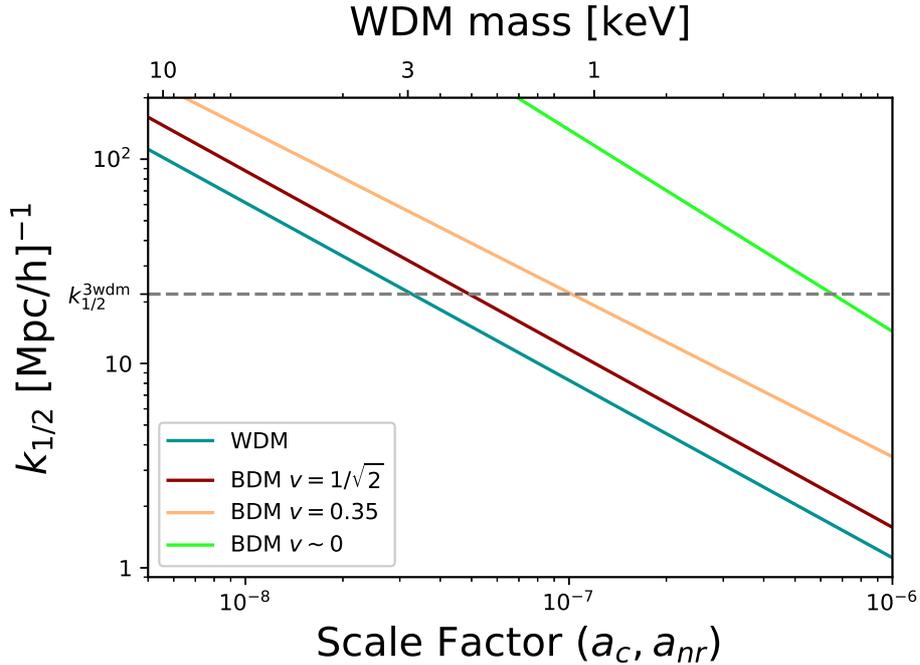}
    \caption{\footnotesize{
       The plot of the $k_{1/2}$ value as a function of $a_c$ in the case of BDM, and the mass for WDM. We assume that the time when the WDM stop being relativistic, $a_{\rm nr}$ is the same as $a_c$. The value of $a_{nr}$ is computed when the WDM momentum $p$ is equal to its mass, $m$, this implies that $v_{\rm wdm}^2 = 1/2$ and therefore $\omega_{\rm wdm} \sim 1/6$. The dotted line is the $k_{1/2}$ value for the 3 keV WDM. }}    \label{fig:k_half}
  \end{minipage}
\end{figure*}

Constraints on the BDM mass can be computed using Eq.(\ref{eq:k_half}), In Figure \ref{fig:k_half} we show lines that fit the numerical values of $k_{1/2}$ obtained from the numerical code CLASS, their value depends on $a_c$ for the BDM case, and the mass, $m_{\rm wdm}$, for the WDM case. 
 
\section{Press-Schecter}\label{ssec:preschechter}

\begin{figure*}
\centering
  \begin{minipage}[t]{.75\textwidth}
      \includegraphics[width=\textwidth]{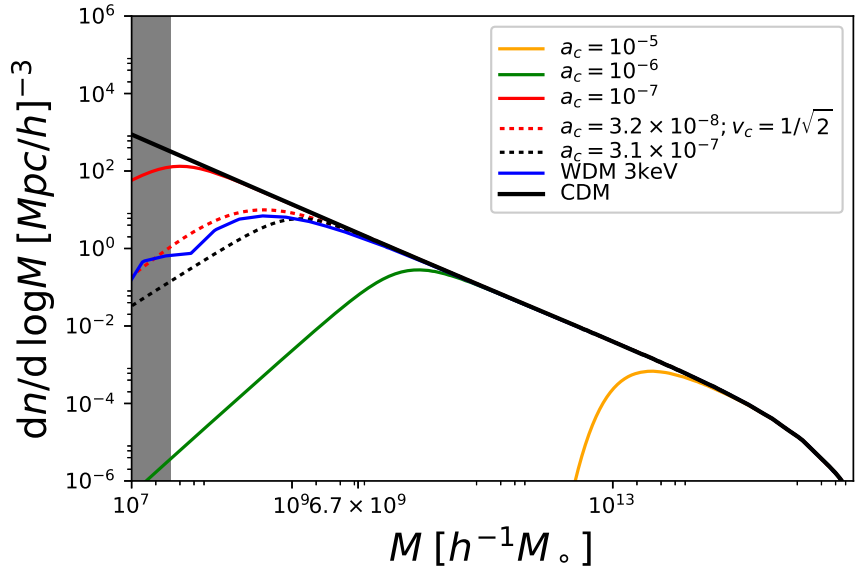}
    \caption{\footnotesize{
        Halo mass function as a function of mass for redshift $z=0$. Lines show the theoretical predictions from the Press-Schechter approach. The black solid line is the CDM model; Blue line is WDM. Other color lines denote results for different BDM models. The yellow, green and red straight lines are BDM models with $v_c = 0$ and $a_c = \{ 10^{-5}, 10^{-6}, 10^{-7}\}$, respectively. Dotted red and black lines are the BDM models that preserve the free-streaming scale obtained from WDM.  }  }  \label{fig:pressschechter}
  \end{minipage}
\end{figure*}

The change in the matter power spectrum is known to strongly affect large scale structure, we include the effects of the abundance of structure in the BDM cosmological model, we adopt the PressSchechter (PS) approach \cite{Press:1973iz}. First, we compute the linear matter power spectrum for the BDM, as described above, and  compute the halo mass function as
\begin{equation}
  \frac{dn}{d\log M} = M \frac{dm}{dM} = \frac{1}{2} \frac{\overline{\rho}}{M} \mathcal{F}(\nu) \frac{d \log \sigma^2}{ d \log M}
\end{equation}
where $n$ is the number density of haloes, $M$ the halo mass and the the peak-height of perturbations is
\begin{eqnarray}
   \nu = \frac{\delta_c^2(z) }{\sigma^2(M)}, 
\end{eqnarray}
where  $\delta_c(z) = \frac{1.686}{D(z)}$ is the overdensity required for spherical collapse model at redshift z in a $\Lambda$CDM cosmology and and $D(z)$ is the linear theory growth function. The evolution of $\delta_{c}(z)$ and $D(z)$ evolve accordingly to the perturbation formalism for BDM introduced in Section \ref{sec:framework}. The average density $\overline{\rho} = \Omega_m \rho_c$ where $\rho_c$ is the critical density of the Universe. Here $\Omega_m = \Omega_{c} + \Omega_b $. The variance of the linear density field  on mass-scale, $\sigma^2(M)$, can be computed from the following integrals
\begin{equation} \label{eq:sigma_8}
  	\sigma^2(M) = \int_0^\infty dk \frac{k^2 P_{\rm lin}(k)}{2\pi^2}| W(kR) |^2
\end{equation}
we use the sharp-k window function $W(x) = \Theta( 1 - kR)$, with $\Theta$ being a Heaviside step function, and $R = (3cM/4\pi \overline{\rho})^{1/3}$, where the value of $c = 2.5$ is proved to be best for cases similar as the WDM \cite{Benson:2012su}. The sharp--k window function has also been prove to better work on models that show cut-off scale al large scales.
Finally for the  first crossing distribution $\mathcal{F}(\nu)$ we adopt \cite{Bond:1990iw}, that has the form
\begin{equation}
   \mathcal{F}(\nu) = A\left( 1 + \frac{1}{\nu^{\prime p}} \right)\sqrt{ \frac{\nu^\prime}{2\pi} } e^{-\nu^\prime/2}
\end{equation}\label{eq:pressschechter}
with $\nu^{\prime} = 0.707\nu$, $p = 0.3$, and $A = 0.322$ determined from the integral constraint $\int f(\nu)d\nu = 1$.   

For mass-scales $M < M_{\rm fs}$, free-streaming erases all peaks in the initial density field, and hence peak theory should tell us that there are no haloes below this mass scale, therefore, significant numbers of haloes below the cut-off mass are suppressed.

We show this behavior more schematically in Figure \ref{fig:pressschechter}, where we compare CDM and BDM mass functions. For large halo masses $M > 10^{13} M_{\odot  } h^{-1}$ the models are indistinguishable for a BDM particle with early transition. However, for smaller halo masses and late transitions, we can see significant suppression in the number of structures.

To compute the value of $f\sigma_8$ first we compute $\sigma_8$ with Eq.(\ref{eq:sigma_8}) for $R = 8$ Mpc/h. For the sake of comparison with previous results, we adopt the top-hat window function to compute $\sigma_8$, using the sharp-k window function we obtained the same behavior but with a 58\% difference respect the top-hat window function. As mention before the spherical top-hat window filter
is not perfect for a truncated power spectrum \cite{Benson:2012su}, but is a conservative choice that would result in weaker bound on the model.

\begin{figure}[t]
  \centering
    \includegraphics[width=0.8\textwidth]{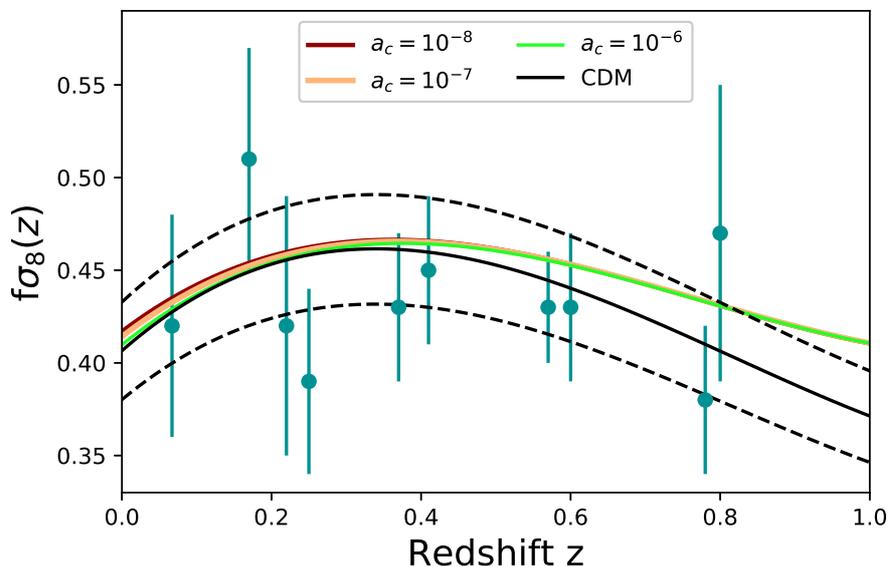}
    \caption{\footnotesize{
        Plots for $f\sigma_{8}$ for CDM (Solid black line) and its $1\sigma$ tolerance error (dashed lines). We also plot the predicted f$\sigma_8$ for the BDM model for different values of $a_c = \{10^{-8}, 10^{-7}, 10^{-6} \}$, red, orange and green lines, respectively. All BDM curves assume $v_c = 0$, other values of $v_c$ have no significant difference between curves. For larger redshift is clearly that both models deviates from each other, this in principle could be the clearest test distinguish between different models. }}
    \label{fig:fs8}
\end{figure}

The growth rate of structure, $f$, is well defined by
\begin{equation}
f \equiv \frac { \mathrm { d } \ln \delta _ { \mathrm { m } } } { \mathrm { d } \ln a }
\end{equation}
The growth rate of structure can be approximated by the parametrization $f = \Omega _ { \mathrm { m } } ( a ) ^ { \gamma }$, where $\gamma$ is commonly referred to as growth index, which is approximately a constant in the range of observations. The definition of the parameter $\Omega_{ \mathrm { m } } (a) \equiv \rho_{ \mathrm { m } } (a) / 3 M_{ P }^{ 2 }H^{ 2 }( a )$ and $\rho_m$ is the density of matter evolution. For the BDM case 
\begin{equation}
   f  = \Omega_m^{0.58}
\end{equation}
for $z<1$, in contrast with the value from $\Lambda$CDM which has an $\alpha = 6/11 \sim 0.545$. Because the evolution of the perturbations. In Figure\ref{fig:fs8} we plot $f\sigma_8$ for CDM and BDM models. We obtain the $1\sigma$ tolerance for $f\sigma_8$ from Montecarlo simulations, we compare this curves with the ones obtained for BDM for different values of $a_c$, we notice that the velocity parameter, $v_c$, has no physical implications on the value of $f\sigma_8$ and it is important to notice that BDM and CDM deviates from one another at large redshift. This could be important for future observations.

\begin{table}[t]
    \centering 
    \caption{$f\sigma_8$ table data.}
    
\begin{tabular}{ p{1.3cm}  p{2.5cm}    p{1.3cm}   p{3.5cm} } 
\hline
z & $f\sigma_8$(z) & 1/k &        Reference \\
\hline
0.067 & $0.42 \pm 0.05$   & 16.0-30 &    6dFGRS(2012) \cite{Beutler:2012px} \\
0.17   & $0.51\pm0.06$     & 6.7-50   &    2dFGRS(2004) \cite{Percival:2004fs} \\
0.22   & $0.42\pm0.07$     & 3.3-50   &    WiggleZ(2011) \cite{Jennings:2010uv} \\
0.25   & $0.35\pm0.06$     & 30-200  &    SDSS \cite{Samushia:2011cs} \\
0.37   & $0.46\pm0.04$     & 30-200  &    SDSS \cite{Samushia:2011cs} \\
0.41   & $0.45\pm0.04$     & 3.3-50   &    2dFGRS(2004) \cite{Percival:2004fs} \\
0.57   & $0.462\pm0.041$ & 25-130  &    BOSS \cite{Alam:2015qta} \\
0.6     & $0.43\pm0.04$     & 3.3-50   &    WiggleZ(2011) \cite{Jennings:2010uv} \\
0.78   & $0.38\pm0.04$     & 3.3-50   &    WiggleZ(2011) \cite{Jennings:2010uv} \\
0.8     & $0.47\pm0.08$     & 6.0-35   &    VIPERS(2013) \cite{delaTorre:2013rpa} \\ 
\hline
\end{tabular}
\label{table2}
\end{table}

\section{Halo Density Profiles}\label{ssec:bdmprofile}


There are essentially two types of profiles, the ones stemming from cosmological $N$-body simulations that have a cusp in its inner region, e.g. Navarro-Frenk-White (NFW) profile \cite{Navarro:1995iw,Navarro:1996gj}. On the other hand, the phenomenological motived cored profiles, such as the Burkert or Pseudo-Isothermal (ISO) profiles \cite{Burkert:1995yz, vanAlbada:1984js}. Cuspy and cored profiles can both be fitted to most galaxy rotation curves, but with a marked preference for a cored inner region with constant density \cite{deBlok:2009sp, delaMacorra:2011df}

As mention before, BDM particles at high densities are relativistic, the galaxy central regions could concentrate high amount of dark matter that BDM could behave as HDM, while in the outer galaxy regions BDM will behave as a non-relativistic particle, this is, as a standard CDM. To come forward this idea in galaxies we introduce a core radius ($r_{\rm core}$) stemming from the relativistic nature of the BDM, besides the scale length ($r_s$) and core density ($\rho_{\rm core}$) typical halo parameters. The galactic core density is going to be proportional to energy of the transition $\rho_{\rm core} \propto \Lambda_c^4$ and the profile properties determine the energy scale of the particle physics model.

The average energy densities in galactic halos is of the order $\rho_g \sim 10^{5} \rho_{\rm cr}$ ($\rho_{\rm  cr}$, being the critical Universe's background density) and  as long as $\rho_g<\rho_{\rm core}$ we expect a standard CDM galaxy profile (given by the NFW profile, $\rho_{{\rm nfw}}$). The NFW profile has a cuspy inner region with $\rho_{{\rm nfw}}$ diverging in the center of the galaxy. Therefore, once one approaches the center of the galaxy the energy density increases in the NFW profile and once it reaches the point $\rho_g=\rho_{\rm core}$  we encounter the BDM phase transition. Therefore, inside $r<r_{\rm core}$ the BDM particles are relativistic and the DM energy density $\rho_{\rm core}$ remains constant avoiding a galactic cusp. Of course we would expect a smooth  transition region between these two distinct behaviors but we expect the effect of the thickness of this transition region to be small and we will not consider it here.

Since our BDM behaves as CDM for $\rho < \rho_{\rm core}$  we expect to have a NFW type of profile. Therefore, the  BDM profile  is assume to be \cite{delaMacorra:2009yb}
\begin{eqnarray} \label{eq:rhobdm}
	\rho_{bdm}  = \frac{2 \rho_{\rm core}}{\left( 1+\frac{r}{r_{\rm core}}\right)\left( 1+\frac{r}{r_s}\right)^2},
	\label{eq:fix-BDM}
\end{eqnarray}
with $r_{\rm core}< r_s$ The BDM profile coincides with $\rho_{{\rm nfw}}$ at large radius but has a core inner region, then we can find a conection with NFW parameters. When the galaxy energy density $\rho_{\rm bdm}$ reaches the value $\rho_{\rm core}=\Lambda_c^4$ at  $r\simeq r_{\rm core} $  and for $r_{\rm core} \ll r_s$ we have
\begin{equation} \label{eq:relation ec_free_bdm_params}
 \rho_{core} = \frac{\rho_0 r_s}{2r_{\rm core}},
\end{equation}
where $r_s$ and $\rho_0$ are typical NFW  halo parameters. The value of $r_{c}$  The parameters $r_{\rm core}$, $\rho_{\rm core}$ and $r_s$ had been estimated fitting galaxy rotation curves \cite{delaMacorra:2011df, Mastache:2011cn}.

\section{Discussion and Conclusions}\label{sec:conclusion}

We have presented the BDM model which introduce an abrupt evolution of the velocity dispersion parameter, $v_c$ as one of the main characteristic in the particle model. Along with the time when the particle became non-relativistic $a_c$, as part of the two parameters that aims to understand the nature of the dark matter. The velocity introduce different effects to the CMB power spectrum, linear matter power spectrum, dark matter halo density profile,  halo mass function and growth rate of structure in the case of BDM. 

The effect of introducing a non-trivial initial velocity dispersion, $v_c$, at the time of transition, $a_c$, prevent clustering inside the Jeans length. We perform the analysis by constraining our model using the 2018 Planck CMB, BAO measurements, and the JLA SNe Ia data in the MonteCarlo run. The Montecarlo constrain the $a_c - v_c$ parameter space establishing the valid region at 1$\sigma$ and 2$\sigma$, this set the lower limits to $a_c$. For the extreme case, where $v_c = 0$, the transition is $a_c \leq 2.66 \times 10^{-6}$ at 1$\sigma$. We also find that the relation between $a_c$ and $v_c$ should preserve the free-streaming for a BDM particle, for which we can find an equivalence to the mass of a WDM particle of $m_{\rm wdm}> 535.5$ eV at $1\sigma$ likelihood. 

We take the 3 keV WDM as the model to compare, although the lower mass for WDM is still unsettle. We find that this particle become non-relativistic at $a_{nr} = 3.18 \times 10^{-8}$, have a free streaming scale $\lambda_{fs}  = 0.10$ Mpc/h, and a Jeans mass of $M_{fs} = 1.97 \times 10^7 \, {\rm M_\odot/h^3}$. The BDM model reproduce the results obtained with WDM if we take at least $v_c = 1/\sqrt{2}$ and the value of the transition is the same as WDM, $a_c = a_{nr}$, with this only condition the $k_{1/2}$ for BDM is only 20\% larger than the one obtained with 3 keV WDM. The interesting result came if we assume that $\lambda_{fs}^{\rm wdm} = \lambda_{fs}^{\rm bdm}$ and force $v_c = 0$, this is, we consider the most abrupt transition in the BDM model. For this case we found that a BDM with a transition at $a_c = 3.18 \times 10^{-7}$ and $v_c = 0$ has the same free-streaming, preserve the cutoff in the matter power spectrum, but its transition is an order of magnitud bigger. If a WDM have this transition it would correspond to a WDM mass of $m_{\rm wdm} = 535.5$ eV.

Although the constrains found for WDM seems to be low, is reasonable because the only reason that CMB is constraining the $a_c$ transition is due to the matter-radiation equality epoch, $a_{eq}$, since this values cannot significantly change because the power spectrum observations, therefore the time of the transition has a lower limit which is related to preserve the amount of radiation at the time of equality. The constrains we found for WDM is of he order of magnitude a previous results: based on the abundance of redshift $z = 6$ galaxies in the Hubble Frontier Fields put constrains of $m_{\rm wdm} > 2.4$ keV \cite{Menci:2016eui}. Based on the galaxy luminosity function at $z \sim 6 - 8$ put constrains on $m_{\rm wdm} > 1.5$ keV  \cite{Corasaniti:2016epp}. Lensing surveys such as CLASH provide $m_{\rm wdm} > 0.9$ keV lower bounds \cite{Pacucci:2013jfa}. The highest lower limit is given by the high redshift Ly-$\alpha$ forest data which put lower bounds of $m_{\rm wdm} > 3.3 keV$  \cite{Viel:2013apy}.  

This framework where we include the dispersion velocity of the dark matter particle may be incorporated in a broad number of observational cosmological probes, including forecasts for large scale structure measures, i.e. weak lensing  \cite{Markovic:2010te}, future galaxy clustering two-point function measures of the power spectrum \cite{vandenBosch:2003nk}. Future observation from large to small-scale clustering of dark and baryonic matter may be able to clear the nature of dark matter and its primordial origin.

\appendix

\section{Appendix  \\  \\Linear Perturbations Equations}\label{appendix:perturbation}

\begin{figure}[t]
  \centering
    \includegraphics[width=\textwidth]{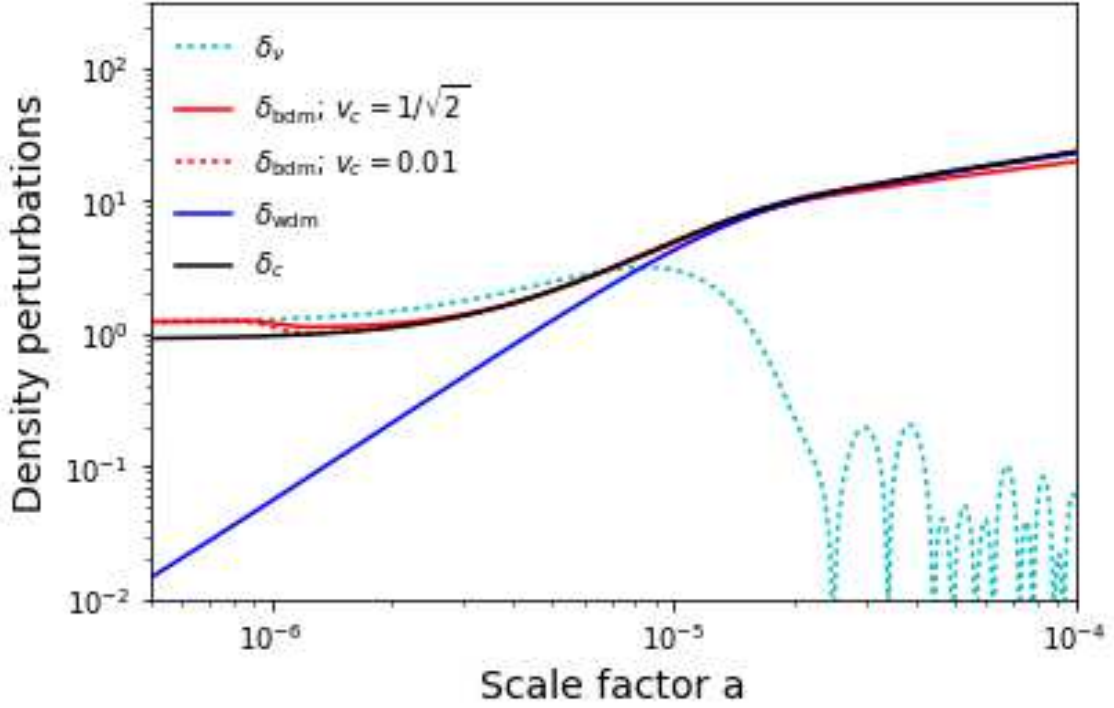}
    \caption{\footnotesize{
        The evolution of the perturbations for CDM ($\delta_c$, black line), a 3 keV mass WDM ($\delta_{\rm wdm}$, blue line), massless neutrino ($\delta_\nu$, dotted green line) as a function of the scale factor, and BDM ($\delta_{\rm bdm}$, both red lines) given by Eqs.(\ref{eq:delta_rad}) and (\ref{delta_mat}). We can distinct a small difference between two cases for BDM, both of them have a transition when $a_c = 10^{-6}$ but the solid red line is when $v_c = 1/\sqrt{2}$, and the dotted red line when $v_c \sim 0$. All perturbations has a wavenumber $k = 1$/Mpc. }}  \label{fig:pert_evol}       
\end{figure}

In this appendix we show the first order equations of the perturbations of the BDM. We follow \cite{Ma:1995ey} and compute the fluid approximation to the perturbed equations in k-space in the synchronous gauge for our BDM model. Before the transition, $a<a_c$, the perturbed equations are:
\begin{eqnarray}\label{eq:BDM_rad}
    \dot{\delta} &=& -\frac{4}{3}\left( \theta + \frac{\dot {h}}{2} \right) \label{eq:delta_rad} \\ 
    \dot{\theta} &=& \frac{1}{4}k^{2}\delta - k^2\sigma, \\
    \dot { \sigma } &=& - 3 \frac{\sigma}{\tau} + \frac{1}{3} \left( 2\theta + \dot{h}  \right) \, ,
\end{eqnarray}

where $\sigma$ is the anisotropic stress perturbations. The dot represent the derivative respect to the conformal time $\tau \equiv \int d t/a(t)$, $H$ is the Hubble parameter, $H \equiv \dot{a}/a$, and $\theta$ is the velocity of the perturbation. Until this point the behavior of the BDM particles are similar as the ultra-relativistic massless neutrinos \cite{Ma:1995ey}.

After the transition, $a>a_c$, the BDM particles goes through the transition. Using Eqs.(\ref{eq:eos_bdm}) and (\ref{eq:rho_bdm})  we are able to compute the perturbation equations:

\begin{align}
    \dot{\delta_c} &= -\left(1+\omega\right)\left(\theta + \frac{\dot{h}}{2} \right) -  \frac{4\omega(1-\omega)}{1+\omega} H\delta_c \label{delta_mat} \\[1ex]
    \dot{\theta} &= -H\theta \frac{ (1 - \omega)(1-3\omega) }{ 1+\omega } + k^2\delta \, \frac{\omega(5-3\omega)  }{ 3(1+\omega)^2} - k^2\sigma  \\[1ex]
    \dot { \sigma } &=
           - 3 \left( \frac { 1 } { \tau } +\frac { 2 H} { 3 } \left[ \frac{1-3\omega}{1+3\omega}  \right] \right) \sigma + 
           \frac { 4 } { 9 } \left( 2 \theta + \dot { h } \right) \frac{\omega (5 - 3\omega)}{  \left( 1+\omega \right)^2} 	  \label{eq:sigma_matter}
\end{align}

In Eq.(\ref{eq:sigma_matter}) we have taken the anisotropic stress approximation for massive neutrinos \cite{Hu:1998kj} and ignore the $\dot{\eta}$ term that slightly improve the computation of the matter power spectrum \cite{Lesgourgues:2011rh}. We have also used the relation ${\dot \omega} = -2 H \omega$. The perturbation evolution for different components of the Universe is shown is Figure \ref{fig:pert_evol} as a function of the scale factor. When $v_c$ takes values $v_c < 1/\sqrt{2}$ the EoS is a non-continuos function, as well as $\delta_{\rm bdm}$, $\theta_{\rm bdm}$ and $\sigma_{\rm bdm}$, therefore has no good numerical solution to the set of equation that describe the perturbation evolution. To overcome this problem we implement a step function in order to smooth the transition and compute a solution for the perturbation.

From Figure \ref{fig:pert_evol} we notice that BDM perturbation behaves always as radiation at early times, the evolution is similar to the massless neutrinos, after the transition start behaving as CDM and only after matter-radiation equality BDM, CDM and WDM (with a mass of 3 keV) has the same behavior.

\acknowledgments

 J. Mastache  acknowledges the supported by CONACyT though Catedras program.  A. de la Macorra  Project IN103518 PAPIIT-UNAM  and  PASPA-DGAPA, UNAM.

\bibliographystyle{unsrtnat}
\bibliography{cmbbdm}

\end{document}